\documentclass[runningheads]{llncs}
\usepackage{amsmath}
\usepackage{amsfonts}
\usepackage{amssymb}
\usepackage{subfigure}
\usepackage[T1]{fontenc}
\usepackage{newtx}
\usepackage{graphicx}
\begin{document}
\title{Self-supervised Spatio-Temporal Graph Mask-Passing Attention Network for Perceptual Importance Prediction of Multi-point Tactility}
\titlerunning{SSTGMPAN for Perceptual Importance Prediction of Multi-point Tactility}
%
\author{Dazhong He\orcidID{0009-0006-5090-2178} \and
Qian Liu
}
%
\authorrunning{D. He and Q. Liu et al.}
%
\institute{Dalian University of Technology, Dalian 116024, China\\
\email{hedazhongcs@mail.dlut.edu.cn qianliu@dlut.edu.cn}}
%
\maketitle              
\begin{abstract}
While visual and auditory information are prevalent in modern multimedia systems, haptic interaction, e.g., tactile and kinesthetic interaction, provides a unique form of human perception. However, multimedia technology for contact interaction is less mature than non-contact multimedia technologies and requires further development. Specialized haptic media technologies, requiring low latency and bitrates, are essential to enable haptic interaction, necessitating haptic information compression. Existing vibrotactile signal compression methods, based on the perceptual model, do not consider the characteristics of fused tactile perception at multiple spatially distributed interaction points. In fact, differences in tactile perceptual importance are not limited to conventional frequency and time domains, but also encompass differences in the spatial locations on the skin unique to tactile perception. For the most frequently used tactile information, vibrotactile texture perception, we have developed a model to predict its perceptual importance at multiple points, based on self-supervised learning and Spatio-Temporal Graph Neural Network. Current experimental results indicate that this model can effectively predict the perceptual importance of various points in multi-point tactile perception scenarios.

\end{abstract}
\section{Introduction}
\begin{sloppypar}
Haptic multimedia technology, crucial for immersive experiences, is still developing~\cite{ref_1}. It provides haptic feedback through capturing, compressing, transmitting, and displaying haptic information. This work focuses on the compression of tactile information. Although single-point tactile interaction data is currently small compared to audio and video data, the future will likely see a shift towards multi-point interaction scenarios. Tens of interaction points achieve a data rate of Mbit/s~\cite{ref_4}, but this is still not enough for high fidelity. In order to meet the needs of realistic large-area contact perception of human skin surface, the number of interaction points can reach thousands~\cite{ref_2}. Increasing the number of interaction points leads to a sharp increase in data rate, making tactile data compression a clear necessity~\cite{ref_4}. Spatial compression algorithms are needed in multi-point scenarios~\cite{ref_1}.
\end{sloppypar}

Existing tactile codecs like PVC-LSP~\cite{ref_5} and VC-PWQ~\cite{ref_4} use psychophysical perceptual models to maintain perception quality at a low data rate. However, existing perceptual models only consider frequency domain perceptual characteristics for single interaction point~\cite{ref_2}, overlooking multi-point spatially distributed interactions. In fact, the masking effect also occurs in the time domain and in the spatial domain between different interaction points in multi-point tactile interactions. Therefore, developing perceptual models for multi-point tactile scenes is a promising direction, as it can guide compression encoding, communication resource scheduling, and other related applications. The challenge in developing multipoint tactile perceptual models is how to analyze perceptual masking over time, frequency, and spatial domains of multiple interaction points. In this work, we specifically focus on fine roughness (which gives rise to vibrotactile perception and is exemplified in the hand) and its perceptual model.

The multi-point perceptual model can be implemented in the form of inputting a multivariate time series of tactile signals formed by each interaction point and outputting the importance index of each. The ``\textbf{Importance Index}" represents perceptual criticality as a classification probability, abstracting this into a multivariate time series classification problem. Interaction points form a graph structure in multi-point tactile interaction, with nodes characterized by tactile time series and connected through hidden dependencies. This making it natural to view the problem from a graph perspective. Graph neural networks (GNNs) have made some achievements in processing graph data. They propagate information within the graph, allowing each node to learn about its surroundings. Particularly, spatio-temporal graph neural networks (STGNNs), handling spatio-temporal data modeling tasks, provide future predictions or category labels based on input time series and optional external graph structure, making them effective for constructing multi-point tactile perceptual models.

\begin{sloppypar}
In this paper, we propose a deep learning model, the \textbf{S}elf-supervised \textbf{S}patio-\textbf{T}emporal \textbf{G}raph \textbf{M}ask-\textbf{P}assing \textbf{A}ttention \textbf{N}etwork (SSTGMPAN), to predict perceptual importance based on vibrotactile time series signals as inputs from each node in multi-point interaction. It uses a novel temporal-spectral mask-passing attention mechanism for capturing the temporal-spectral masking dynamics between nodes and self-supervised learning for improved data utilization and model generalization, reducing the need for extensive manual labeling.
\end{sloppypar}

\section{Related Work}
\subsection{Vibrotactile Perceptual Model and Its Applications}
Human tactile perception involves sensing body surfaces and perceiving external stimuli, including friction, hardness, temperature, and roughness~\cite{ref_3}. Surface roughness perception depends on roughness fineness. Macroroughness involves visible height profiles on the object surface and its regularity~\cite{ref_1}. Fine roughness, or microscopic roughness, results from high-frequency vibration when body parts slide on the object surface.

In single-point tactile perception, temporal and frequency masking occur~\cite{ref_2}. Temporal masking reduces sensitivity over time due to a strong stimulus, while frequency masking raises the threshold around a dominant frequency. In multi-point tactile perception, spatial masking occurs, where stimulation at one interaction point increases the perception threshold at surrounding points~\cite{ref_13}. Existing vibrotactile codecs like VC-PWQ~\cite{ref_4} and PVC-SLP~\cite{ref_5} apply vibrotactile stimulation sensing characteristics to vibrotactile signal compression, focusing on frequency domain masking and absolute perception thresholds~\cite{ref_2}. Multi-point tactile coding necessitates the introduction of time and spatial domain masking for advanced data compression.

\subsection{Spatio-Temporal Graph Neural Network}
\begin{sloppypar}
Traditional GNN struggle with dynamic spatio-temporal data. The rise of spatio-temporal graphs, with evolving nodes and dynamic relationships, has led to Spatio-temporal Graph Neural Networks. These networks integrate temporal convolutional or recurrent layers with graph convolutional operations to capture spatial and temporal dynamic dependencies. Initially, STGNNs were used for forecasting, particularly in traffic flow. However, they have recently been applied to other tasks for time series, e.g., classification~\cite{ref_27}, demonstrating their ability to model complex spatio-temporal dependencies.
\end{sloppypar}

\subsection{Self-supervised learning}
Self-supervised learning (SSL) is a deep learning paradigm that learns from unlabeled data, leveraging the data's inherent structure or statistical properties. It's successful in CV and NLP, and divided into generative and contrastive methods. Generative SSL has advanced in graph learning, with GraphMAE~\cite{ref_50} proposing a graph-oriented MAE for graph data reconstruction after masking.

\section{Methodology and Model}
\begin{figure}[t]
	\centering
	\includegraphics[width=1.0\textwidth]{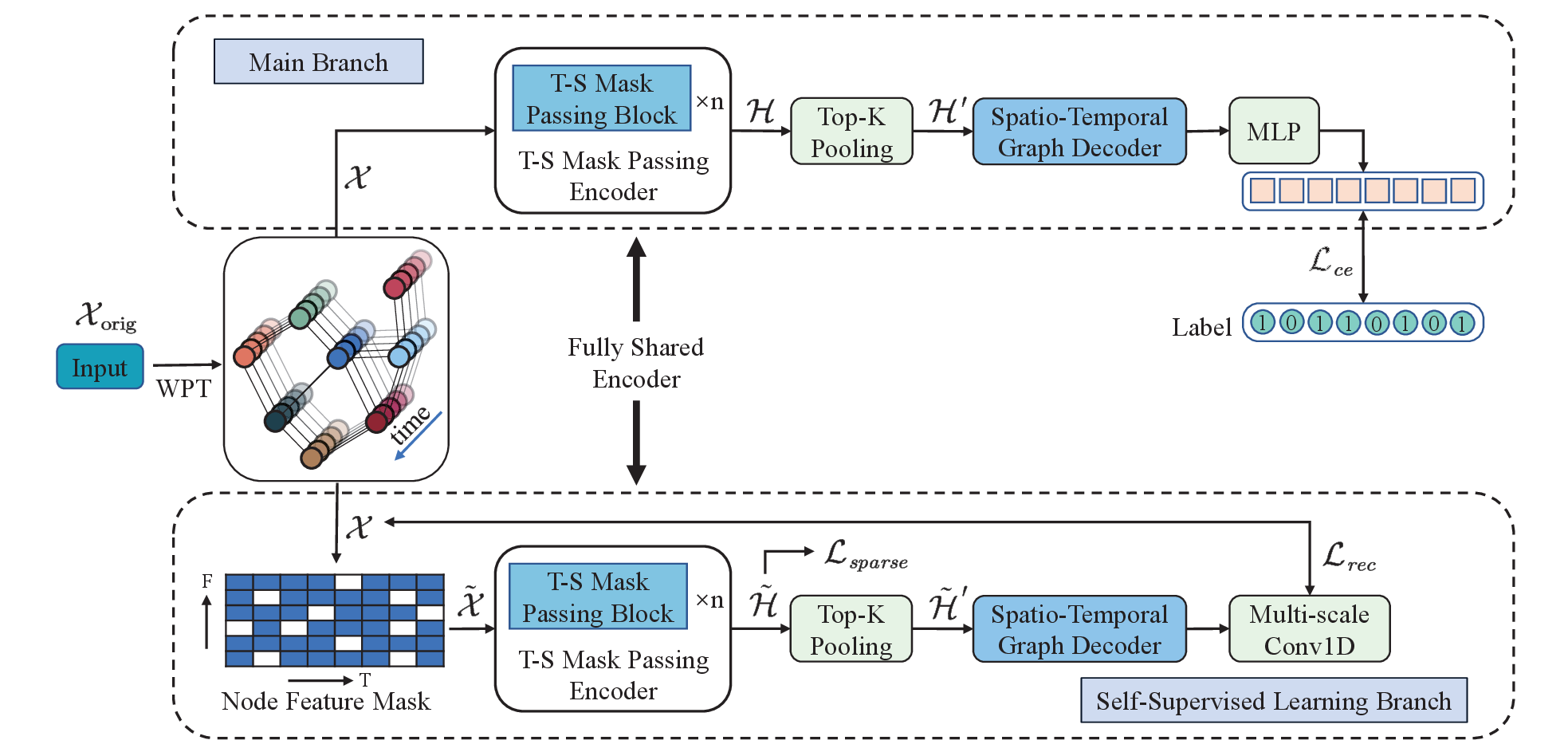}
	\caption{The overall architecture of SSTGMPAN.} \label{model architecture}
\end{figure}

In this study, we propose a new spatio-temporal graph neural network for multi-point tactile interaction scenes. It analyzes the masking relationship of multi-point vibrotactile perception in the time, frequency, and spatial domains, and predicts the perceptual importance of interaction points. The technical details of our model are elaborated in this section, and its architecture is shown in Fig.~\ref{model architecture}.

\subsection{Preliminaries and Problem Formulation}
The vibrotactile signal at $N$ points is divided into $F$ frequency bands over $T$ time steps, represented as $\mathcal{X}^{(t-T+1):t} \in \mathbb{R}^{N\times F\times T}$. This allows the signal to be segmented into short time slices in order to accommodate changes in perceptual importance over time. We aim to predict $\hat{\mathcal{Y}} \in [0, 1]^{N}$, where $\hat{\mathcal{Y}}_k$ represents probability that the $k^{th}$ series is key to tactile perception. 
We give formal definitions below.

\begin{definition}
	Graph: A graph is represented by $\mathcal{G}=(\mathcal{V},\mathcal{E})$, where $\mathcal{V}$ is the set of nodes for tactile interaction and $\mathcal{E}$ is the set of edges, with $|\mathcal{V}|=N$ nodes.
\end{definition}

\begin{definition}
    Node Neighborhood: The neighborhood of a node $v_i$ is defined as $\mathcal{N}(v_i)=\{v_j\in \mathcal{V}|(v_j,v_i)\in \mathcal{E}\}$.
\end{definition}

\begin{definition}
	Vibrotactile Signals: The vibrotactile signal matrix $\mathcal{X}^{(t-T+1):t} = \{x_1, x_2, \\ \cdots, x_N \} \in \mathbb{R}^{N\times F\times T}$ represents each node's characteristics in the graph. For a specific node $v_j\in \mathcal{V}$, its vibrotactile signal is $x_j\in \mathbb{R}^{F\times T}$.
\end{definition}

\begin{definition}
	Perceptual Importance Labels: For each node $i$, the label $\mathcal{Y}_i\in\{0,1\}$ indicates its perceptual importance, i.e., whether the node is a key node for reproducing human tactile perception. If there are $N$ nodes, the labels form an N-dimensional vector $\mathcal{Y}=[\mathcal{Y}_1,\mathcal{Y}_2,\ldots,\mathcal{Y}_N]$.
\end{definition}
\textbf{Problem statement:} We construct a function mapping $f(\cdot)$ by minimizing the binary cross-entropy loss. This predicts $\hat{\mathcal{Y}}$ based on the input $\mathcal{X}^{(t-T+1):t}$ and $\mathcal{G}$, where $\hat{\mathcal{Y}}_i \in [0, 1]$ for $i = 1, 2, .. ., N$. The relation is represented as follows:
\begin{equation}
[\mathcal{X}^{(t-T+1):t},\mathcal{G}]\xrightarrow{f}\hat{\mathcal{Y}},
\end{equation}
where $\mathcal{X}^{(t-T+1):t}=\{x_1, x_2, \cdots, x_N \} \in \mathbb{R}^{N\times F\times T}$ and $\hat{\mathcal{Y}}\in [0,1]^N$.

\subsection{Architecture and Pipeline of SSTGMPAN}
The top-level idea of the model is to use graph SSL methods to leverage limited training data and improve output representation. The model is trained on pretext and downstream tasks simultaneously, with the pretext task acting as a regularization of the downstream tasks~\cite{ref_52}. Our model has two branches: one for multivariate time series classification, and another for generative self-supervised learning. The two branches share a common encoder, i.e., a temporal-spectral mask-passing encoder. The SSL branch is essentially a masked autoencoder~\cite{ref_40}. 

\begin{figure}[t]
	\begin{center}
		\includegraphics[width=0.6\textwidth]{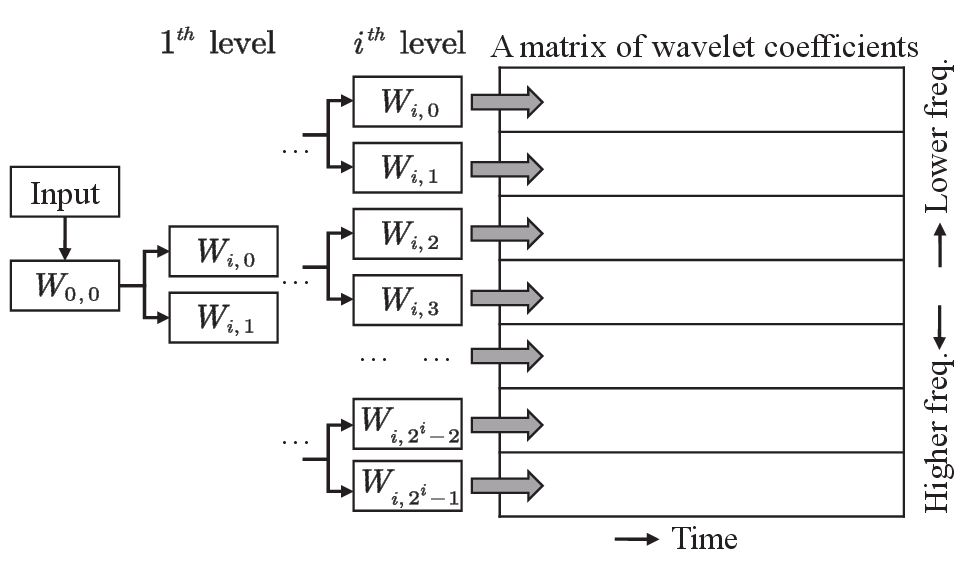}
		\caption{The process of wavelet packet decomposition.} \label{wpt}
	\end{center}
\end{figure}

We use wavelet packet transform (WPT) as a pre-processing step to convert the original time series data $\mathcal{X}_{\text{orig}} \in \mathbb{R}^{N\times T_{\text{orig}}}$ into a new feature representation $\mathcal{X} \in \mathbb{R}^{N\times F\times T}$. Wavelet packet decomposition, a discrete algorithm for analyzing non-stationary signals, decomposes the signal into several frequency bands with wavelet packet coefficients in each. The process is shown in Fig.~\ref{wpt}.

The main branch takes $\mathcal{X}$ and $\mathcal{G}$, generates a latent representation, and uses a multilayer perceptron (MLP) for node classification, trained with cross-entropy loss $\mathcal{L}_{ce}$. We formalize this as:

\begin{equation}
\theta^*,\phi^*=\arg\min_{\theta,\phi}\mathcal{L}_{ce}\Big(MLP\big(p_\phi\big(f_\theta(\mathcal{ G},\mathcal{X})\big)\big),\mathcal{Y}\Big),
\end{equation}
where $f_\theta$ and $p_\phi$ are the encoder and decoder, parameterized by $\theta$ and $\phi$, respectively.

We randomly mask patches in $\mathcal{X}$, and get a corrupted input $\tilde{\mathcal{X}}$ for self-supervised learning. The SSL branch's exclusive decoder fulfills the pretext task, i.e., generating feature representations for input reconstruction. Multiscale depthwise convolution is used for final feature reconstruction. In addition, applying custom sparse constraints to the encoder output eliminates redundant information and extracts key features. This constraint, along with the feature reconstruction constraint, mutually confine each other within the SSL branch, making feature reconstruction challenging and enhancing the model's representational capacity. The SSL branch is formalized as follows:
\begin{equation}
\theta^*,\phi^*_{ssl}=\begin{cases}
    \arg\min_{\theta,\phi_{ssl}}\mathcal{L}_{rec}\Big(Conv\big(p_{\phi_{ssl}} \big(f_\theta(\mathcal{G},\tilde{\mathcal{X}})\big)\big),\mathcal{X}\Big) \\ 
    \arg\min_{\phi}\mathcal{L}_{sparse}\big(f_\theta(\mathcal{G},\tilde{\mathcal{X}})\big)
\end{cases},
\end{equation}
where $f_\theta$ is the shared encoder, and $p_{\phi_{ssl}}$ is the branch-specific decoder.

In a multi-task learning framework, the model's two branches learn jointly. The main branch uses cross-entropy loss for the classification task, while the SSL branch uses reconstruction and sparsity-constraint losses as regularization to mitigate overfitting. The reconstruction loss, a combination of cosine similarity error and mean square error (MSE), aids in comprehensive feature learning and reconstruction. The sparsity-constraint loss encourages the model's encoder to acquire sparser latent representations. These losses collectively form the objective for model training. The shared encoder, under the influence of the main branch, identifies critical nodes and extracts their spatio-temporal masking relationships, During joint learning, it adjusts its sparse features to match the degraded features produced by the temporal-spectral masking. This process degrades each node's features to varying degrees, aligning with sparsity. Through both branches' collaborative action, the encoder learns from the complex data of the spatio-temporal graph, creating a sparse latent representation that retains key information and carries interaction information between nodes.

\subsection{Graph Temporal-Spectral Mask-Passing Attention}
In this subsection, we will introduce a novel temporal-spectral mask-passing attention mechanism, which is the core of the model, responsible for extracting the temporal-spectral masking relationship between nodes. We use an attention mechanism to adaptively capture the dynamic mutual influence correlations between nodes in spectral and temporal dimensions.

We treat any node $v_i \in \mathcal{V}$ as a central node that receives masking influence information from its neighborhood $\mathcal{N}(v_i)$. For the node $v_i$ and any node in its neighborhood $v_j \in \mathcal{N}(v_i)$, they have the feature $x_i\in \mathbb{R}^{F\times T}$, $x_j\in \mathbb{R}^{F\times T}$. 

We first compute their spectral representations $\mathbf{F}_i$ and $\mathbf{F}_j  \in \mathbb{R}^{F\times T}$, and their temporal representations $\mathbf{T}_i$ and $\mathbf{T}_j \in \mathbb{R}^{T \times F}$. They are defined as:
\begin{align}
\mathbf{F}_i = \mathbf{x}_i \mathbf{W}_{Q_{F}}, &\quad \mathbf{F}_j = \mathbf{x}_j \mathbf{W}_{K_{F}} \\
\mathbf{T}_i = \mathbf{x}_i \mathbf{W}_{Q_{T}}, &\quad \mathbf{T}_j = \mathbf{x}_j \mathbf{W}_{K_{T}}
\end{align}
where $\mathbf{W}_{Q_{F}}, \mathbf{W}_{K_{F}} \in \mathbb{R}^{T \times T}$ and $\mathbf{W}_{Q_{T}}, \mathbf{W}_{K_{T}} \in \mathbb{R}^{F \times F}$ are learnable weights, with $\mathbf{W}_{Q_{F}}$ lower triangular and $\mathbf{W}_{K_{F}}$ upper triangular to ensure causality. Each column $\mathbf{F}_i[:,t]$ in the spectrum representation $\mathbf{F}_i$ indicates that the neighbor effect of node $i$ at time step $t$ only impacts subsequent time steps. Each column $\mathbf{F}_j[:,t]$ in the spectrum representation $\mathbf{F}_j$ indicates the influence of node $j$ on other nodes at time step $t$ is the combined effect of all previous time steps. Next, we define attention coefficients $\alpha_{F_{ij}}$ and $\alpha_{T_{ij}}$ as follows:
\begin{equation}
\alpha_{Fij} = \mathrm{softmax} \left( \frac{\mathbf{F}_i \mathbf{W}_F \mathbf{F}_j^{\top}}{\sqrt{d_{T}}} \right), 
\alpha_{Tij} = \mathrm{softmax} \left( \frac{\mathbf{T}_j \mathbf{W}_T \mathbf{T}_i^{\top}}{\sqrt{d_{F}}} \right)
\end{equation}
where $\alpha_{Fij}$ represents the spectral attention between nodes $i$ and $j$, and $\alpha_{Tij}$ represents the temporal attention between nodes $i$ and $j$. To enhance the representation ability, the learnable matrices $\mathbf{W}_F$ and $\mathbf{W}_T$ are added, where $\mathbf{W}_F$ is a lower triangular matrix to ensure that node $ i$ at a specific time step $t$ is only affected by node $j$ before time step $t$. For the same purpose, when calculating $\alpha_{Tij}$, a mask operation precedes the softmax operation.

\begin{figure}[tbp]
	\begin{center}
		\begin{minipage}[b]{\textwidth}
			\centering
			\subfigure[]{
				\includegraphics[width=0.45\textwidth]{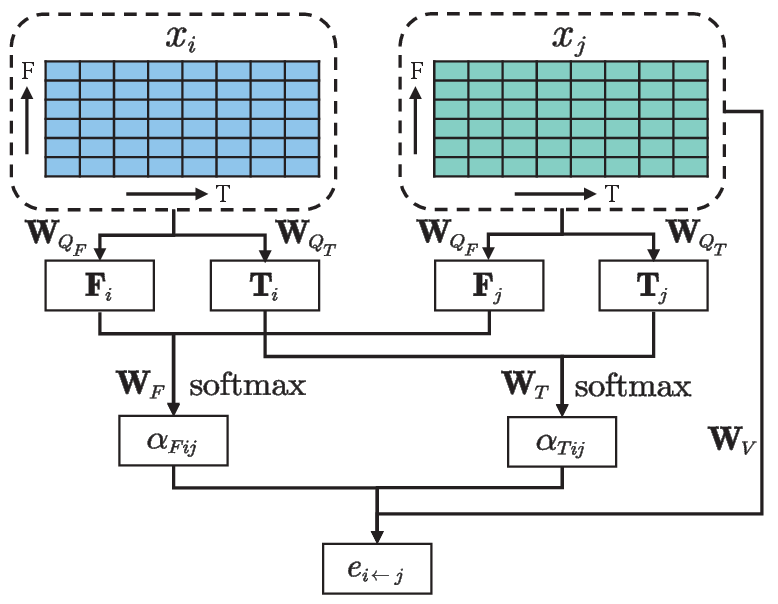}
				\label{stmp-a}
				}
			\subfigure[]{
				\includegraphics[width=0.4\textwidth]{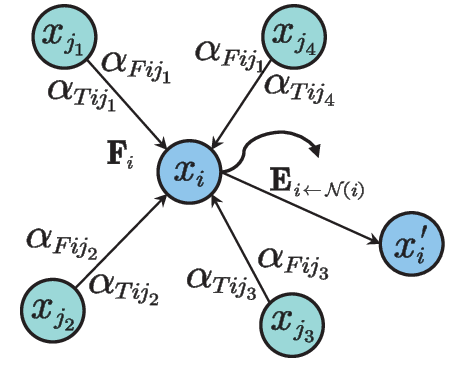}
				\label{stmp-b}
				}
		\end{minipage}
		\caption{Temporal-spectral mask-passing. (a) Calculating the effect of node $j$ on node $i$. (b) Aggregating node neighbors' effects and updating node $i$'s representation.}
		\label{stmp}
	\end{center}
\end{figure}

We formulate a mask-passing strategy that involves weighted aggregation of node features considering both spectral and temporal dimensions. To stabilize learning process, we use multiple heads of attention and integrate their output using 1$\times$1 convolution, delaying the application of the final nonlinearity. Finally, the mask-passing operator is defined as:
\begin{equation}
\mathbf{E}_{i \gets\mathcal{N}(i)} = \sigma \left( Conv_{1\times1} \left( \underset{k-1}{\overset{K}{\Vert}}   \sum_{j \in \mathcal{N}(i)}  \alpha_{Fij} \cdot (\mathbf{x}_j \odot \mathbf{W}_V) \cdot \alpha_{Tij} \right) \right).
\end{equation}
where $\mathbf{W}_V \in \mathbb{R}^{F \times T}$ is a learnable weight matrix, and $\sigma$ is a non-linear activation function. The output $\mathbf{E}_{i \gets\mathcal{N}(i)}$ is element-wise multiplied with the features of node $i$, represented as $\mathbf{x}_i$, yielding the final output features $\mathbf{x}_i^{\prime}$ for each node:
\begin{equation}
\mathbf{x}_i^{\prime} = \mathbf{x}_i \odot \mathbf{E}_{i \gets\mathcal{N}(i)}.
\end{equation}

\subsection{Global Spatial Attention}
Inspired by~\cite{ref_22}, we define a global spatial attention mechanism as an attention matrix $\mathbf{S} \in \mathbb{R}^{N \times N}$, representing node association strength and capturing spatial correlation. The attention matrix is computed as follows:
\begin{equation}
\mathbf{S}=\mathbf{V}_{s}\cdot\sigma(
({\mathcal X}_{h}^{(r-1)}\mathbf{W}_{1})\mathbf{W}_{2}(\mathbf{W}_{3}{\mathcal X}_{h}^{(r-1)})^{T} \\
+ (({\mathcal X}_{h}^{(r-1)})^{T}\mathbf{U}_{1})\mathbf{U}_{2}(\mathbf{U}_{3}{\mathcal X}_{h}^{(r-1)})
+ \mathbf{b}_{s}),
\end{equation}
\begin{equation}
\mathbf{S}_{i,j}^{\prime}=\frac{\exp(\mathbf{S_{i,j}})}{\sum_{j=1}^{N}\exp(\mathbf{S_{i,j}})},
\end{equation}
where ${\mathcal X}_{h}^{(r-1)} \in \mathbb{R}^{N \times F_{r-1} \times T}$ is the input of the $r^{th}$ decoder building block, and $F_{r-1}$ is the spectral feature dimension in the $r^{th}$ layer (when $r=1$, $F_0=F$). $\mathbf{V}_{s}$, $\mathbf{b}_{s}\in\mathbb{R}^{N\times N}$, $\mathbf{W}_{1} \in \mathbb{R}^T$, $\mathbf{W}_{2}\in\mathbb{R}^{F_{r-1}\times T}$, $\mathbf{W}_{3} \in \mathbb{R}^{F_{r-1}}$, $\mathbf{U}_{1}\in\mathbb{R}^{F_{r-1}}$, $\mathbf{U}_{2}\in\mathbb{R}^{T \times F_{r-1}}$, $\mathbf{U}_{3}\in\mathbb{R}^{T}$ is some learnable weights, and $\sigma$ is a sigmoid function.

\subsection{Main Branch for Node Classification}
The main branch, shown in Fig.~\ref{model architecture}, predicts node probabilities to identify critical nodes for vibrotactile perception. We will introduce its main components below.

\subsubsection{Temporal-Spectral Mask-Passing Encoder}
\begin{sloppypar}
The encoder, adopting the proposed temporal-spectral mask-passing attention and fed with $\mathcal{X} \in \mathbb{R}^{N \times F \times T}$ including sinusoidal positional encoding~\cite{ref_55}, aims to embed complex and highly dynamic masking relationships between nodes into a latent representation.
\end{sloppypar}

\subsubsection{Top-K Feature Selection Module}
The Top-K feature selection module enhances model performance by distilling  highly influential key features from the input matrix $\mathcal{H} \in \mathbb{R}^{N \times F \times T}$, resulting in $\mathcal{H}^{\prime} \in \mathbb{R}^{N \times F \times T}$. It ranks features by absolute magnitude, retaining the top percentile.

\subsubsection{Spatio-temporal Graph Decoder}
\begin{figure}[t]
	\begin{center}
		\begin{minipage}[b]{\textwidth}
			\centering
			\subfigure[]{
				\includegraphics[width=0.42\textwidth]{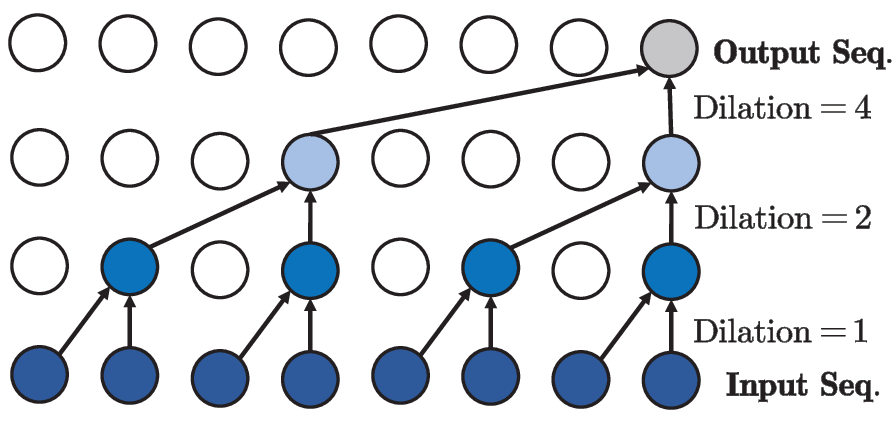}
				\label{decoder-a}
				}
			\subfigure[]{
				\includegraphics[width=0.52\textwidth]{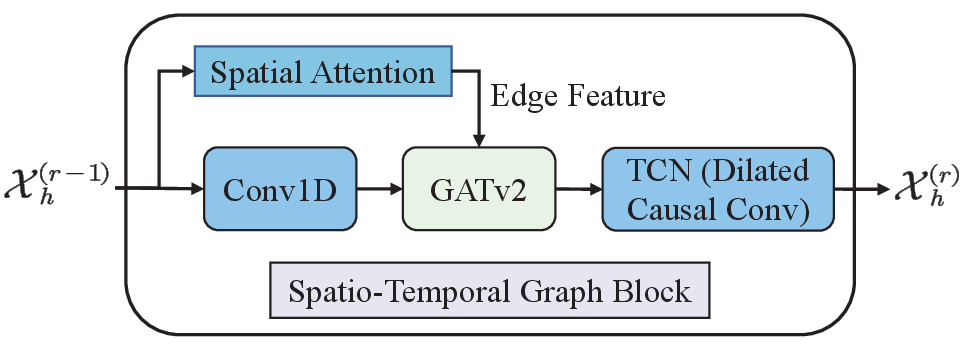}
				\label{decoder-b}
				}
		\end{minipage}
		\caption{(a) Dilated causal convolution. (b) The building block for decoder.}
		\label{decoder}
	\end{center}
\end{figure}

The structure of spatio-temporal graph decoder is shown in Fig.~\ref{decoder-b}. We employ TCN~\cite{ref_56} as temporal convolutional layer to capture the temporal trend of nodes. It uses dilated causal convolution to exponentially expand the receptive field and maintain temporal order through zero padding. GATv2~\cite{ref_58} is used to model node spatial associations. Furthermore, $\mathbf{S}$, the edge feature from global spatial attention, is inputted into GATv2.

\subsection{Self-supervised Learning Branch}
The structure of SSL branch is basically consistent with the main branch, with the addition of a masking strategy and sparse constraints.

We generate a mask matrix $\mathcal{M}$ via random sampling, and elements are set to 0 (default is 1) at a probability of $\alpha_m$. The masked input $\tilde{\mathcal{X}}=\mathcal{X}\odot\mathcal{M}$ is assigned to the SSL branch, and the original $\mathcal{X}$ is used as the reconstruction target.

A custom sparse constraint, inspired by~\cite{ref_60}, is applied to the shared encoder's output to remove redundant information. The constraint loss is defined as the ratio of $L_1$ and $L_p$ norms, with a lower bound to prevent excessive sparsity. For the encoder's output $\tilde{\mathcal{H}} \in \mathbb{R}^{N\times F\times T}$, the sparse constraint loss is formalized as:
\begin{equation}
\mathcal{L}_{sparse}(\tilde{\mathcal{H}})=\frac{l_1\left(\tilde{\mathcal{H}} \right)/n}{l_p\left(\tilde{\mathcal{H}} \right)/n^{1/p}}=n^{1/p-1}\frac{l_1(\mathbf{x})}{l_p(\mathbf{x})} \in [n^{1/p-1}, 1],
\end{equation}
where $n$ is the total number of elements in $\mathcal{H}$. Particularly, $p$ is 4 in this paper.

\section{Experiments}

\subsection{Data Collection and Processing}
To validate our approach, we constructed a dataset towards the perceptual characteristics of multi-point tactile perception, as no such dataset currently exists.

\subsubsection{Multi-point Vibrotactile Signals Acquisition}
\begin{figure}[t]
	\begin{center}
		\begin{minipage}[b]{\textwidth}
			\centering
			\subfigure[]{
				\includegraphics[width=0.31\textwidth]{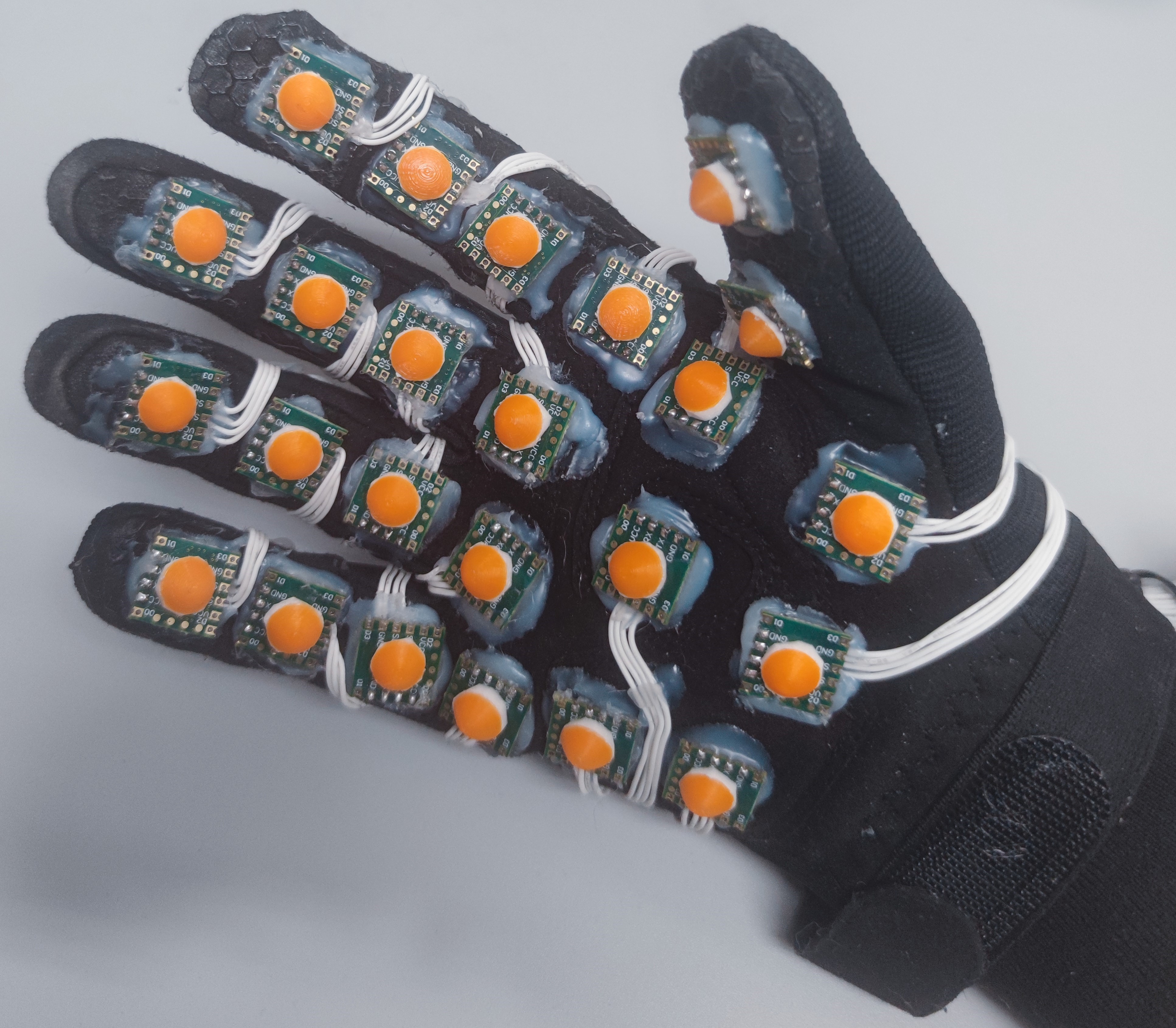}
				\label{fig_gloves}
				}
			\subfigure[]{
				\includegraphics[width=0.36\textwidth]{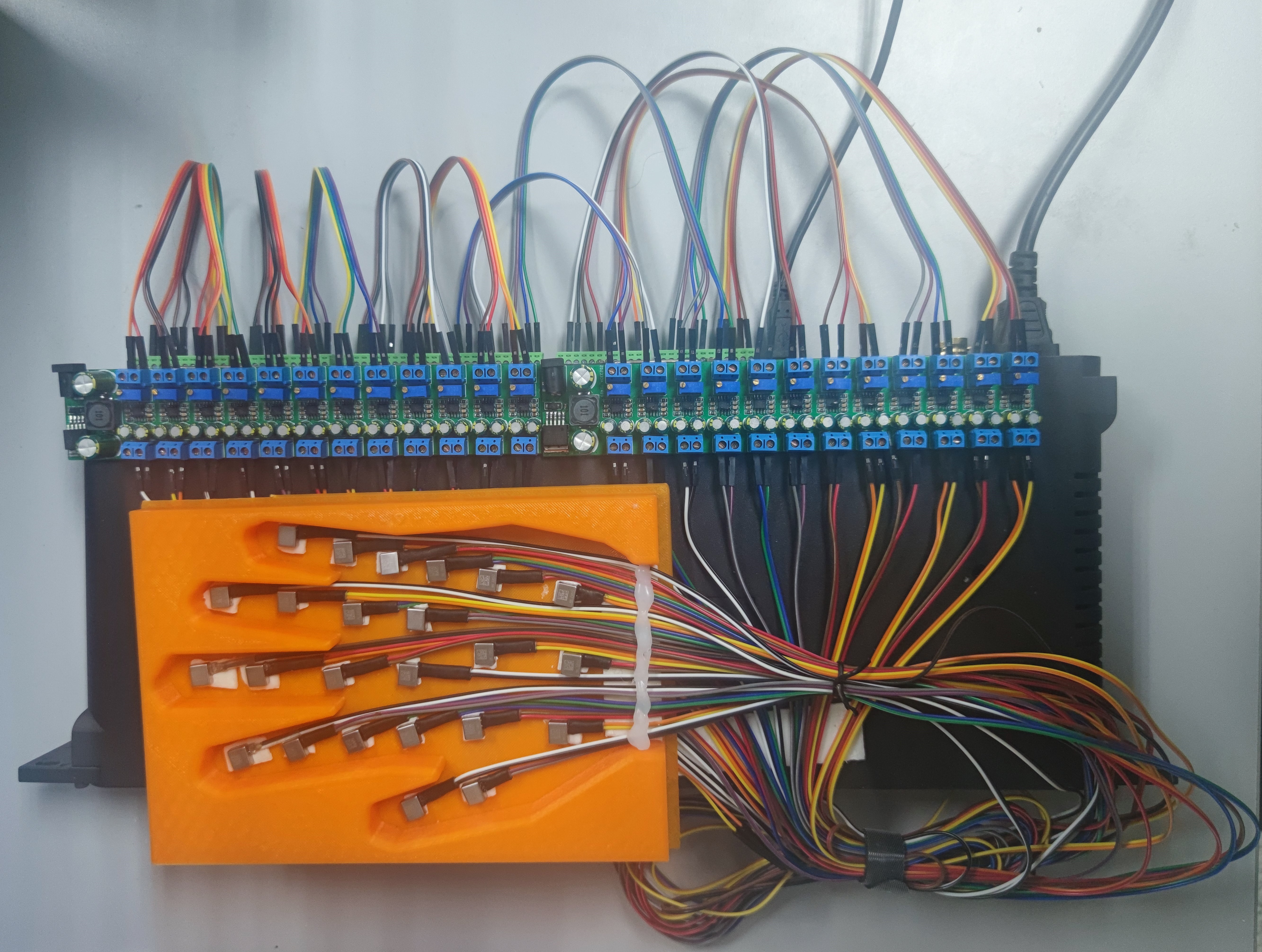}
				\label{fig_gloves_for_annotation}
				}
		\end{minipage}
		\caption{The device used for data collection. (a) The glove acquiring multi-point vibrotactile signals. (b) The device are used for human perceptual importance annotation.}
		\label{fig_device}
	\end{center}
\end{figure}

We recorded vibrotactile signals of a hand interacting with a textured surface using a custom 24-sensor data glove, as shown in Fig.~\ref{fig_gloves}. The device uses multiple pose sensors (WT9011G4K from Witmotion Co.) to record 3-axis acceleration during tactile interaction. Sensors are placed on the glove's palm and fingers, with the z-axis perpendicular to the surface. 3D-printed spacers protect the sensors and transmit vibrations from object interaction. The sensors collect triaxial acceleration data at a 2,000 HZ sampling rate. Wearing tactile data gloves, the researcher explores texture surfaces with natural movements. Texture surfaces include textiles, tabletops, paper, tiles, plastics, metals, and more. Concurrently, variations in normal force and scanning speed are introduced by the researcher (by varying the pressing force and movement speed of the hand within a reasonable range), aiming to observe the system's behavior under a diverse range of interaction conditions. The collected three-axis acceleration signals are high-pass filtered at 10 Hz to eliminate gravity and purposeful human movement effects~\cite{ref_62}. The DFT321 algorithm~\cite{ref_63} maps the filtered signal onto a single axis, given humans' inability to discern high-frequency vibration direction~\cite{ref_64}.

\subsubsection{Perceptual Importance Annotation}
Vibrotactile signals are split into non-overlapping 512-length time segments (ca. 256 milliseconds). Maximum absolute value normalization is applied across all signal channels for signal playback ease. The signals are played through a multi-channel audio interface connected to a computer, driving the vibrotactile actuator (LRA ELA0809 from AAC Technologies, F0 = 170 HZ). All subjects were current students from DUT and were clearly informed of the experiment's purpose and potential risks. The subjects' palm and the palmar side of the fingers made direct contact with the actuators, perceiving the vibrotactile effect rendered by the actuators. The subjects were asked to identify and mark points where they distinctly perceived the signal. The total duration and number of samples annotated per participant are strictly limited (\textasciitilde 40 samples per person, lasting approximately 1 hour, with a break every 10 samples) to prevent inaccuracies from muscle fatigue. The device for annotation is shown in Fig.~\ref{fig_gloves_for_annotation}.

\subsection{Experimental Setup}

\subsubsection{Implementation Details}
We split the dataset into training, validation, and test sets (with split ratio 6:2:2). All the experiments were conducted on a server equipped with an NVIDIA RTX 4090 GPU. Our model, implemented in PyTorch, using the Adam optimizer and PCGard~\cite{ref_65} to update all parameters. We set the initial learning rate to 0.0001, halving it every 50 epochs, and trained for 300 epochs with a batch size of 32. The shared encoder and both decoders have 2 layers each. The SSL branch masking rate $\alpha_m$ is 0.1, and the top-k ratio is 0.8.

\subsubsection{Evaluation Metrics and Baseline Methods}
We evaluate our model using Accuracy, AUC, and F1 score, and compare it with several baselines: MLP with four hidden layers, LSTM, ResNet18 with 1D convolution, OS-CNN with all-scale receptive field~\cite{ref_66}, and TodyNet~\cite{ref_67}, a GNN-based time-series classifier. For all methods, we either adapt the original source code minimally to fit our dataset, or implement by ourselves with similar parameter quantity.

\subsection{Experimental Results}


\subsubsection{Comparison with Baseline Methods}

\begin{figure}[t]
	\begin{center}
		\begin{minipage}[b]{\textwidth}
			\centering
			\subfigure[]{
				\includegraphics[width=0.46\textwidth]{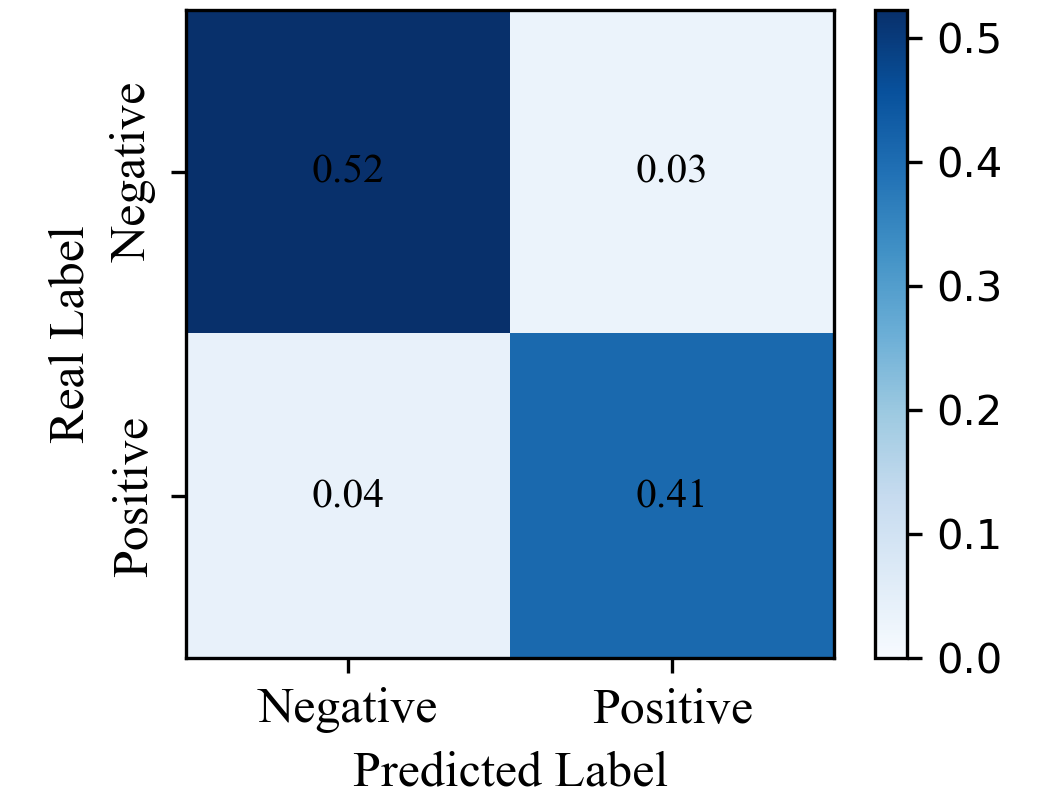}
				\label{confusion_matrix}
				}
			\subfigure[]{
				\includegraphics[width=0.4\textwidth]{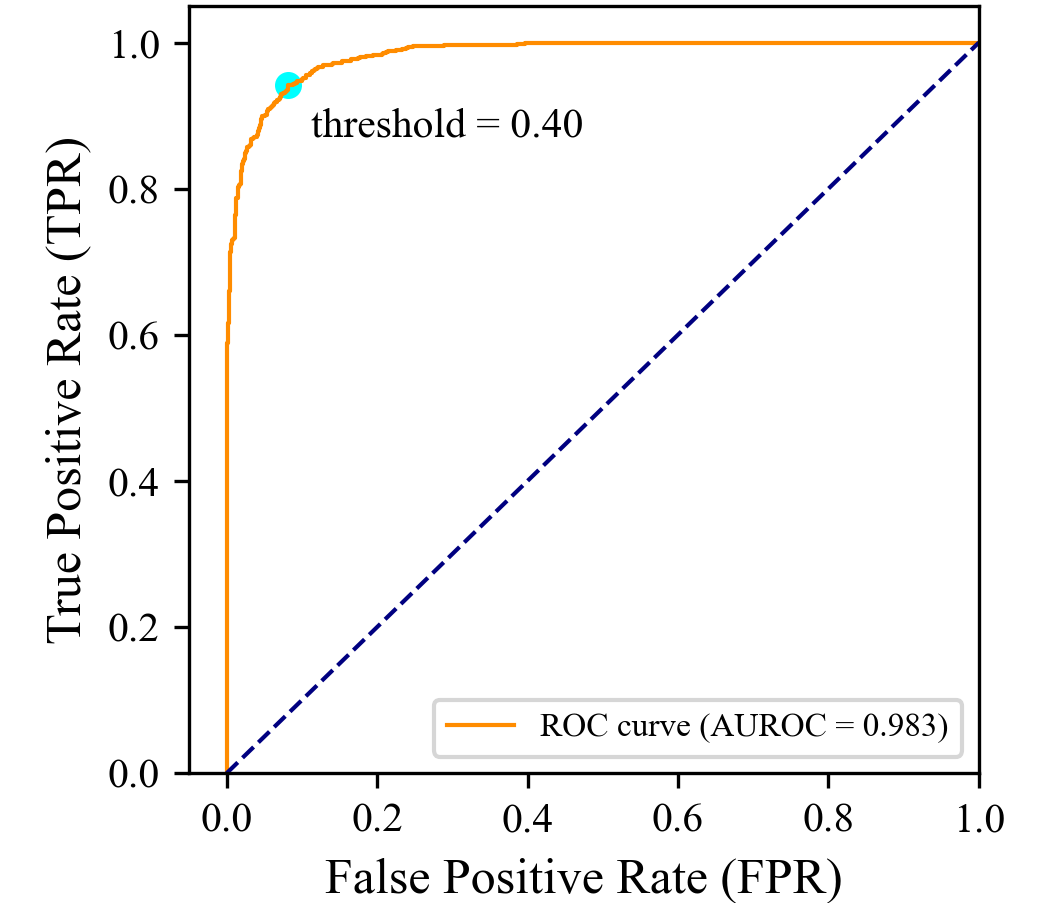}
				\label{ROC_Curve}
				}
		\end{minipage}
		\caption{The results of our model. (a) The confusion matrix. (b) The ROC curve.}
		\label{fig_results}
	\end{center}
\end{figure}

Table~\ref{tab_results1} shows the comparison results of our model with the baseline methods. It can be seen that our model achieves the best performance on all metrics. The ROC curve and confusion matrix of our model are shown in Fig.~\ref{fig_results}. The results demonstrate the effectiveness of our model in predicting the importance of multi-point vibrtactile perception.
\begin{table}[htbp]
	\centering
	\begin{minipage}{0.45\textwidth}
		\centering
		\caption{Classification performance.}\label{tab_results1}
		\begin{tabular}{|c|c|c|c|}
		\hline
		Models      & Acc.   & AUC   & F1 \\
		\hline
		MLP         & 81.65\% & 0.904 & 0.802\\
		LSTM        & 84.79\% & 0.930 & 0.833\\
		ResNet-1D   & 84.73\% & 0.930 & 0.832\\
		OS-CNN      & 88.34\% & 0.956 & 0.875\\
		TodyNet     & 87.32\% & 0.943 & 0.855\\
		SSTGMPAN    & \textbf{92.82\%} & \textbf{0.983} & \textbf{0.939}\\
		\hline
		\end{tabular}
	\end{minipage}
	\begin{minipage}{0.50\textwidth}
		\centering
		\caption{Results for ablation study.}\label{tab_results2}
		\begin{tabular}{|c|c|c|c|}
		\hline
		Model setting   & Acc.   & AUC   & F1 \\
		\hline
		w/o Top-k       & 92.04\% & 0.977 & 0.928\\
		w/o GSA         & 90.87\% & 0.966 & 0.909\\
		w/o SSL         & 89.42\% & 0.962 & 0.891\\
		w/o S-T Decoder & 88.16\% & 0.950 & 0.865\\
		w/o TSMP        & 87.13\% & 0.941 & 0.851\\
		SSTGMPAN        & \textbf{92.82\%} & \textbf{0.983} & \textbf{0.939}\\
		\hline
		\end{tabular}
	\end{minipage}
\end{table}

\subsubsection{Ablation Study}
To verify the effectiveness of each component of our model, we conduct ablation experiments. We denote SSTGMPAN without different components as: w/o Top-k (without the top-k feature selection), w/o GSA (without the global spatial attention), w/o SSL (without the SSL branch, i.e., only the main branch), w/o S-T Decoder (main branch without spatio-temporal graph decoder), and w/o TSMP (without the temporal-spectral mask-passing).

Table~\ref{tab_results2} shows that removing any component significantly reduces our model's performance. The TSMP and the S-T graph decoder are most critical, with their removal decreasing accuracy by 5.69\% and 4.66\%. The SSL branch and the GSA also impact performance, with accuracy drops of 3.40\% and 1.95\% respectively. The sparsity constraint and the top-k feature selection have the least impact.

\subsubsection{Computation Efficiency Comparison}
We compared the training and inference time per 10-batch (with a size of 32) and the number of parameters between our model and the baseline methods, as shown in Table~\ref{tab_results3}. Despite having fewer or similar parameters, our model may take longer in training and inference time, yet it outperforms them in performance. This is primarily due to the GNN in our model, which considers the relationship between nodes, thereby increasing computational complexity. However, this increased complexity allows our model to better capture and understand the inherent structure of the data, leading to improved performance. Therefore, we believe this trade-off between computational speed and model effectiveness is worthwhile. It's also worth noting that the longer training time of ours is attributed to the additional training required for the SSL branch, which is not needed during inference, and thus, is considered acceptable.

\begin{table}[htbp]
	\centering
	\begin{minipage}{1.0\textwidth}
		\centering
		\caption{Computation efficiency comparison.}\label{tab_results3}
		\begin{tabular}{|c|c|c|c|}
		\hline
		Models      & Training time (ms) & Inference time (ms) & \# Params (k)\\
		\hline
		MLP         & 20.2   & 11.8  & 633\\
		LSTM        & 447.0  & 153.1 & 527\\
		ResNet-1D   & 96.1   & 36.2  & 1072\\
		OS-CNN      & 45.0   & 14.6  & 1158\\
		TodyNet     & 709.2  & 207.7 & 502\\
		SSTGMPAN    & 2012.1 & 285.5 & 575\\
		\hline
		\end{tabular}
	\end{minipage}
\end{table}

\subsubsection{Inspection of Class Prototype}
\begin{figure}[htbp]
	\begin{center}
		\begin{minipage}[b]{\textwidth}
			\centering
			\subfigure[]{
				\includegraphics[width=0.45\textwidth]{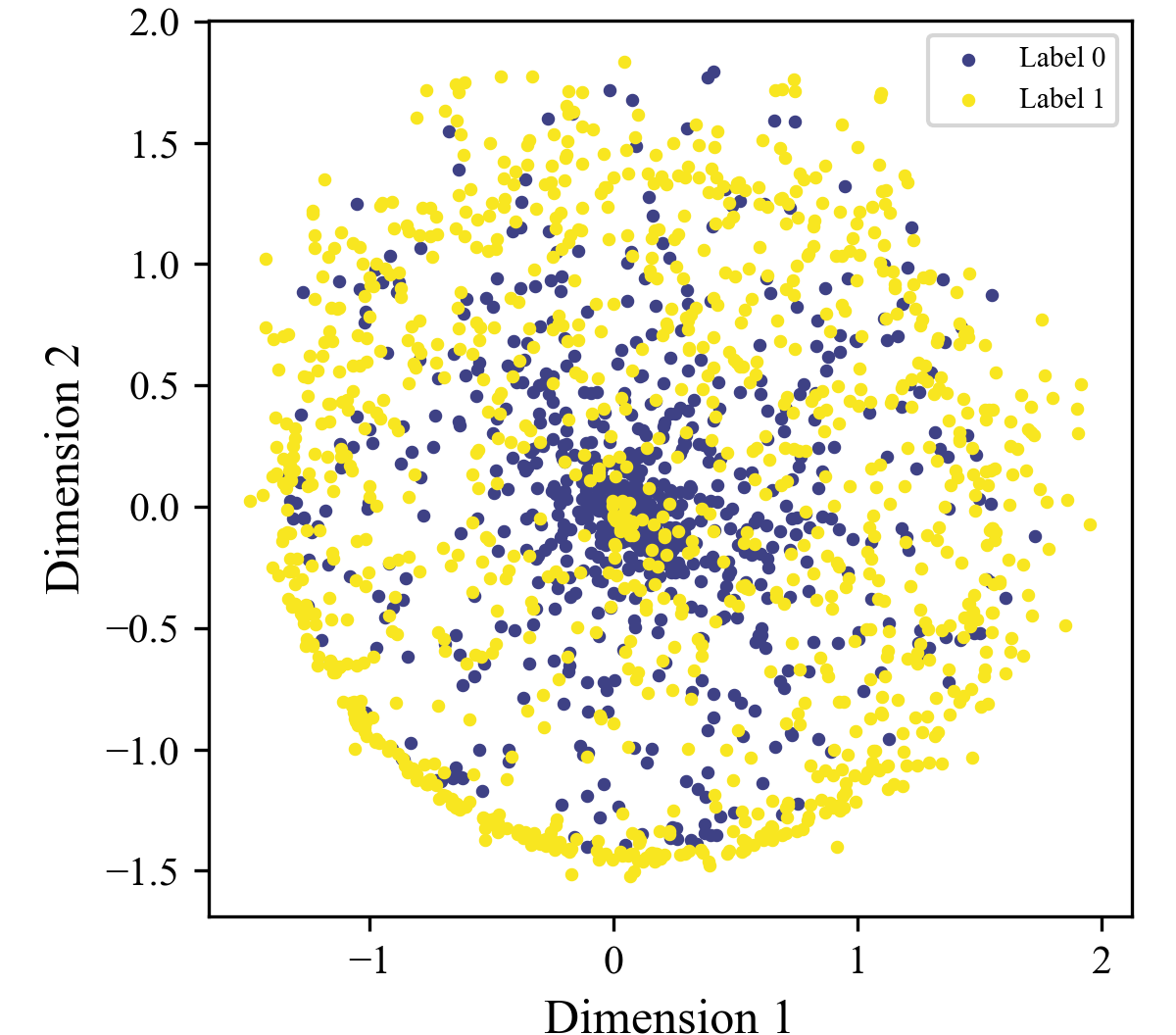}
				\label{fig_tsne_a}
				}
			\subfigure[]{
				\includegraphics[width=0.45\textwidth]{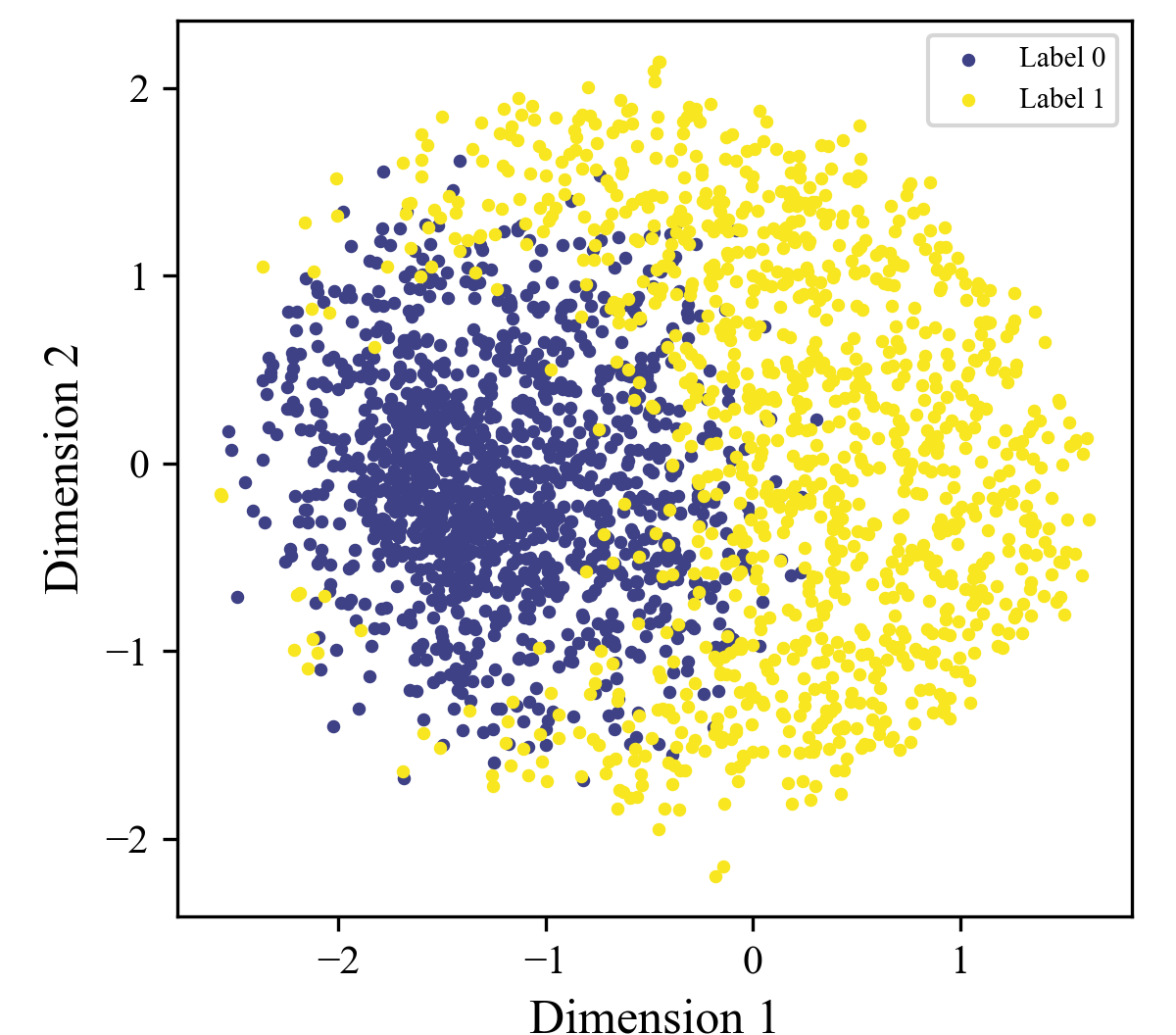}
				\label{fig_tsne_b}
				}
		\end{minipage}
		\caption{The t-SNE visualization. (a) Original. (b) Transformed.}
		\label{fig_tsne}
	\end{center}
\end{figure}
Using t-SNE~\cite{ref_68}, we visualize the main branch's latent representation in Fig.~\ref{fig_tsne}. The close grouping of same-class samples and separation of different classes, indicate that SSTGMPAN effectively characterizes class prototypes for classification.

\section{Conclusion}
In this paper, we propose a novel spatio-temporal graph neural network for perceptual importance prediction in multi-point tactile interaction. Our model uses a proposed temporal-spectral mask-passing attention mechanism to capture dynamic node masking relationships. It outperforms baseline methods. Future work will explore its application in the field of tactile perception, e.g., tactile perception enhancement and compression of tactile perception data.

\begin{credits}
\subsubsection{\ackname} This work was funded by the National Science Foundation of China (Grant No. 62071083), the Liaoning Applied Basic Research Program (Grant No. 2023JH2/101700364), and the Dalian Science and Technology Innovation Foundation (Grant No. 2022JJ12GX014).

\subsubsection{\discintname}
The authors have no competing interests to declare that are
relevant to the content of this article.
\end{credits}
%
%
%
\bibliographystyle{splncs04}
\bibliography{references}
%




\end{document}